\documentclass[10pt,conference]{IEEEtran}
\usepackage{amsmath,amssymb}

\newcommand{\Cc}{{\cal C}}
\newcommand{\xv}{\mathbf{x}}
\newcommand{\phiv}{\boldsymbol{\phi}}
\newcommand{\psiv}{\boldsymbol{\psi}}

\newcommand{\Uv}{\mathbf{U}}
\newcommand{\yv}{\mathbf{y}}
\newcommand{\Yv}{\mathbf{Y}}
\newcommand{\zv}{\mathbf{z}}

\newcommand{\yvh}{\hat{\mathbf{y}}}
\newcommand{\zvh}{\hat{\mathbf{z}}}
\newcommand{\Yvh}{\hat{\mathbf{Y}}}
\newcommand{\Real}{\mathbb{R}}

\newcommand{\e}{\epsilon}

\newcommand{\db}{\bar d}

\def\P{{\mathrm P}}

\newtheorem{theorem}{Theorem}
\newtheorem{lemma}{Lemma}

\begin{document}

\title{Sparse Linear Representation}

\author{
\authorblockN{Halyun Jeong and Young-Han Kim}
\authorblockA{Dept. of ECE, UCSD\\
Email: \{hajeong, yhk\}@ucsd.edu}
}
% \and
% \authorblockN{Muriel Medard}
% \authorblockA{Laboratory for Information and Decision Systems \\
% MIT \\
% Cambridge, MA \\
% Email: medard@mit.edu}
% \and
% \authorblockN{Amin Shokrollahi}
% \authorblockA{Lab. of Math. Algorithms \\
% EPFL\\
% Lausanne, Switzerland \\
% Email: amin.shokrollahi@epfl.ch}
% \and
% \authorblockN{Ram Zamir}
% \authorblockA{EE - Systems Dprt.\\
% Tel Aviv University\\
% Tel Aviv, Israel \\
% Email: zamir@eng.tau.ac.il }
%}
% avoiding spaces at the end of the author lines is not a problem with
% conference papers because we don't use \thanks or \IEEEmembership
%Homer Simpson\authorrefmark{2},
%Montgomery Scott\authorrefmark{3} and
%Eldon Tyrell\authorrefmark{4}}
%Telephone: (800) 555--1212, Fax: (888) 555--1212}
%\authorblockA{\authorrefmark{4}Tyrell Inc., 123 Replicant Street, Los Angeles, California 90210--4321}}

\maketitle

\begin{abstract}
This paper studies the question of how well a signal can be reprsented
by a sparse linear combination of reference signals from an
overcomplete dictionary. When the dictionary size is exponential in
the dimension of signal, then the exact characterization of the
optimal distortion is given as a function of the dictionary size
exponent and the number of reference signals for the linear
representation. Roughly speaking, every signal is sparse if the
dictionary size is exponentially large, no matter how small the
exponent is. Furthermore, an iterative method similar to matching
pursuit that successively finds the best reference signal at each
stage gives asymptotically optimal representations. This method is
essentially equivalent to successive refinement for multiple
descriptions and provides a simple alternative proof of the successive
refinability of white Gaussian sources.
\end{abstract}

\section{Introduction and Main Results}
Suppose one wishes to represent a signal as a linear combination of
reference signals. If the collection $\Cc$ of reference signals
(called {\em dictionary}) is rich (i.e., the size $M = |\Cc|$ of
dictionary is much larger than the dimension $n$ of the signal) or if
one is allowed to take an arbitrarily complex linear combination
(i.e., the number $k$ of reference signals forming the linear
combination is very large), then one can expect that the linear
representation approximates the original signal with very little
distortion. As a trivial example, if $\Cc$ contains $n$ linearly
independent reference signals of dimention $n$, then every signal can
be represented faithfully as a linear combination of those $n$
reference signals. On the other extreme point, if $\Cc$ includes all
possible signals, then the original signal can be represented as (a
linear combination of) itself without any distortion. More generally,
Shannon's rate distortion theory~\cite{Shannon1959} suggests that if
the dictionary size $M = 2^{nR}$ is exponential in $n$ with exponent
$R > 0$, then the best reference signal (as a singleton) achives the
distortion $D(R)$ given as a function of $R$.

Several interesting questions arise:
\begin{enumerate}
\item
What will happen if the linear combination is sparse ($k \ll n$)? How
well can one represent a signal as a (sparse) linear combination of
reference signals?

\item How should one choose the dictionary of reference signals under
the size limitation? Is there a dictionary that provides a good
representation for all or most signals?

\item How can one find the best linear representation given the
dictionary? Is there a low-complexity algorithm with optimal or
near-optimal performance?
\end{enumerate}

These questions arise in many applications and naturally have been
studied in several different contexts~\cite{mallat2009}.
The current paper provides partial
answers to these questions by focusing on asymptotic relationship
between the collection size $M$, the dimension $n$ of the signal, the
sparsity $k$ of the representation, and the distortion $D$ of the
representation.

More formally, let $\Cc = \{\phiv(1), \phiv(2), \ldots, \phiv(M)\}$ be
a collection (dictionary) of $M$ vectors in $\Real^n$. For each vector
$\yv \in \Real^n$, we define its best $k$-linear representation
$\yvh_k$ from the dictionary $\Cc$ as
\[
\yvh_k = x_1 \phiv(m_1) + x_2 \phiv(m_2) + \cdots + x_k \phiv(m_k),
\]
where $x_1,\ldots, x_k \in \Real$ and $m_1, \ldots, m_k \in [1:M] :=
\{1,2,\ldots,M\}$ are chosen to minimize the squared error
\begin{align*}
&d_k(\yv,\Cc)= \|\yv - ( x_1 \phiv(m_1) + x_2 \phiv(m_2) + \cdots + x_k \phiv(m_k) )\|^2.
\end{align*}
Here the norm of a vector $\zv = (z_1,\ldots, z_n) \in\Real^n $ is
defined as $||\zv||= (\sum_{i=1}^n z_i^2)^{1/2}$.

We further define the worst-case distortion $d_k^{*}(\Cc)$ of the
dictionary $\Cc$ as
\[
d^{*}_k(\Cc) := \sup_{\yv: \|\yv\|^2 \le 1} d_k(\yv,\Cc),
\]
where the supremum is taken over all $n$-vectors $\yv$ in the (closed)
unit sphere.

Note that $d_k^{*}(\Cc) \le 1$ for all $\Cc$ and all $n$, with
$d_k^{*}(\Cc) = 1$ attained by a singleton dictionary $\Cc =
\{\boldsymbol{0}\}$. Conversely, if $M < n$, then $d_k^{*}(\Cc) =
1$ for any dictionary $\Cc$ of size $M$. Hence, we consider the case
$M \ge n$ only, that is, the case in which the dictionary is {\em
overcomplete}.

Similarly, we define the average-case distortion $\db_k(\Cc)$ of
the dictionary $\Cc$ as
\[
\db_k(\Cc) = E\left(d_k(\Yv, \Cc)\right),
\]where the expectation is taken with respect to a random signal
$\Yv$ uniformly drawn from the unit sphere $\{\yv \in \Real^n: \|\yv\|
\le 1\}$.

Now we are ready to state our main results. The first result concerns
the existence of an asymptotically good dictionary.
\begin{theorem}
Suppose $M = M_n$ satisfies
\[
\liminf_{n\to\infty} \frac{\log M}{n} > 0.
\]
Then there exists a sequence of dictionaries $\Cc_n$ of respective
sizes $M_n$ such that
\begin{align}
\limsup_{n\to\infty}
\bigg [ \log d_k^{*}(\Cc_n) + \frac{2k \log M}{n} \bigg ] \le 0. \label{eq:thm1}
\end{align}
In particular, if $k \to \infty$ as $n \to \infty$, then
$d_k^{*}(\Cc_n) \to 0$.
\end{theorem}
\medskip

An interesting implication of Theorem 1 is that if we choose a good
dictionary of exponentially large size, no matter how small the
exponent is, every signal is essentially sparse (say, $k = \log\log
n$) with respect to that dictionary in the asymptotics.

The proof of Theorem 1 will be given in Section II. The major
ingredients of the proof include Wyner's uniform sphere covering
lemma~\cite{Wyner1967} and its application in successive linear
representation. Simply put, given a good dictionary for singleton
representations ($k = 1$), we iteratively represent the signal, the
error, the error of the error, etc.\@ by scaling the same dictionary.

This representation method is intimately related to successive
refinement coding~\cite{Equitz--Cover1991}. Indeed, Theorem~1,
specialized to $k=1$, is essentially equivalent to Shannon's rate
distortion theorem for white Gaussian sources~\cite{Shannon1959}.  At
the same time, the representation method gives a very simple proof of
successive refinability~\cite{Equitz--Cover1991} and additive
successive refinability~\cite{Tuncel--Rose2003} of white Gaussian
sources under the mean squared error distortion.

It turns out that the asymptotic distortion in Theorem 1, which is
achieved by the simple successive representation method, is in fact
optimal. The following result, essentially due to Fletcher
{\it et al.} \cite{Fletcher--Rangan--Goyal--Ramchandran2006},
provides the performance bound for the optimal dictionary.

\begin{theorem} [{\it{\cite[Theorem 2]{Fletcher--Rangan--Goyal--Ramchandran2006}}}]
For any sequence of dictionaries $\Cc_n$ of size $M = M_n$ and any
nondecreasing sequence $k = k_n$,
\begin{align*}
\liminf_{n\to\infty} \bigg [ \log \db_k(\Cc_n) + \frac {2 \log {M\choose k}}{n-k}
 + c_n \bigg ] \ge 0.
\end{align*}
where
\begin{align*}
c_n = \log \frac{n}{n-k} + \frac{k}{n-k}\log\frac{n}{k}.
\end{align*}\\
In particular, if $k$ is bounded,
then for any sequence of dictionaries $\Cc_n$ of size $M = M_n$,
\[
\liminf_{n\to\infty} \bigg [ \log \db_k(\Cc_n) + \frac {2 k\log M}{n-k} \bigg ] \ge 0.
\]
\end{theorem}

\vskip .1in

Note that if $M = 2^{nR}$ for some $R > 0$ and $k$ is a constant, then Theorem 2
implies that the average distortion is lower bounded by
\[
\liminf_{n\to\infty} \bigg [ \log \db_k(\Cc_n) + \frac{2k \log M}{n} \bigg ] \ge 0.
\]
(Therefore so is the worst-case distortion.) Thus the distortion bound
in \eqref{eq:thm1} Theorem 1 is tight when the dictionary size grows
exponentially in $n$.

The asymptotic optimality of successive representation method provides
a theoretical justification for matching
pursuit~\cite{Mallat--Zhang1993} or similar greedy algorithms in
signal processing. This conclusion is especially appealing since these
iterative methods have linear complexity in dictionary size $M$ (or
even lower complexity if the dictionary has further structures), while
finding the optimal representation in a single shot, even when
tractable, can have much higher complexity.  However, there are two caveats.
First, the dictionary size here is exponential in the signal
dimension. Second, the dictionary should represent all signals with
singletons uniformly well.

In a broad context, these results are intimately related to {\em
recovery} of sparse signals via linear measurements. Indeed, the
sparse linear representation can be expressed as
\begin{equation}
\label{eq:system}
\yv = \Phi \xv + \zv,
\end{equation}
where $\Phi$ is an $n \times M$ matrix with columns in $\Cc,$ $\xv \in
\Real^M$ is a sparse vector with $k$ nonzero elements $x_1,\ldots,
x_k$, and $\zv$ is the representation error. The award-winning papers
by Candes and Tao~\cite{Candes--Tao2006} and Donoho~\cite{Donoho2006}
showed that a sparse signal $\xv$ can be reconstructed {\em exactly}
and {\em efficiently} from the measurement $\yv$ given by the
underdetermined system~\eqref{eq:system} of linear equations (when the
measurement noise $\zv = 0$), opening up the exciting field of
compressed sensing. There have been several follow-up discussions that
connect compressed sensing and information theory; we refer the reader
to \cite{Sarvotham--Baron--Baraniuk2006, Aeron--Zhao--Saligrama2006,
Fletcher--Rangan--Goyal2007, Wainwright2007, Zhang--Pfister2008,
Weidai--Hoavinhpham--Olgicamilenkovic2009, Jin--Rao2008a}
for various aspects of the connection between two
fields.

While it is quite unfair to summarize in a single sentence a variety
of problems studied by compressed sensing and more generally sparse
signal recovery, the central focus therein is to recover the {\em
true} sparse signal $\xv$ from the measurement $\yv$. In particular,
when the measurement process is corrupted by noise, the main goal
becomes mapping a noise-corrupted measurement output $\yv$ to its
corresponding cause $\xv$ in an efficient manner.

The sparse signal representation problem is in a sense dual to the
sparse signal recovery problem (just like source coding is dual to
channel coding). Here the focus is on $\yv$ and its representation
(approximation). There is no {\em true} representation, and a good
dictionary should have several alternative representations of similar
distortions. As mentioned above, the problem of general---not
necessarily linear---sparse representation (also called the sparse
approximation problem) has a history of longer than a
century~\cite{Temlyakov2003, Schmidt1907} and has been studied in several
different contexts. Along with the parallel development in compressed
sensing, the recent focus has been efficient algorithms and their
theoretical properties; see, for example, \cite{Tropp2004, Tropp2006,
Donoho--Elad--Temlyakov2006}.

In comparison, studies in~\cite{Fletcher--Rangan--Goyal--Ramchandran2006,Akcakaya--Tarokh2008},
and this paper focus on finding asymptotically optimal dictionaries, regardless of
computational complexity,\footnote{Fortuitously, the associated
representation method is highly efficient.} and study the tradeoff among
the sparsity of the representation, the size of the dictionary, and
the fidelity of the approximation. For example, Fletcher
{\it et al.} \cite{Fletcher--Rangan--Goyal--Ramchandran2006} found
a lower bound on the approximation error using rate distortion theory for
Gaussian sources with mean squared distortion. A similar lower bound is
obtained by Akcakaya and Tarokh~\cite{Akcakaya--Tarokh2008} based on careful
calculation of volumes of spherical caps.
Thus, the main contribution of this paper is twofold. First, our Theorem 1
shows that these lower bounds (in particular the one in
\cite[Theorem 2]{Fletcher--Rangan--Goyal--Ramchandran2006}) are tight in
asymptotic when the dictionary size is exponential in signal length.
Second, we show that a simple successive representation
method achieves the lower bound, revealing an intimate connection between sparse
signal representation and multiple description coding.

The rest of the paper is organized as follows.  We give the proof of
Theorem 1 in Section II. In Section III, we digress a little to
discuss the implication of Theorem 1 on successive refinement for
multiple descriptions of white Gaussian sources and its
dual---successive cancelation for additive white Gaussian noise
multiple access channels. Finally, the proof of Theorem 2 is presented
in Section IV.

\section{Successive Linear Representation}
In this section, we prove that there exists a codebook of exponential
size that is asymptotically good for all signals and all sparsity
level. More constructively, we demonstrate that a simple iterative
representation method finds a good representation.

More precisely, we show that if $R'_0 > R_0 > 0$, there exists a
sequence of dictionaries $\Cc_n$ with sizes $M = 2^{nR'_0}$ such that
for $n = n(R'_0, R_0)$ sufficiently large,
\[
d_k^{*}(\Cc_n) \le 2^{-2kR_0}
\]
for every $k$ (independent of n).
Since the above inequality holds for all $R_0 \in (0, R'_0)$, we have
\[
\limsup_{n\to\infty} \bigg [ \log d_k^{*}(\Cc_n) + \frac{2k \log M}{n} \bigg ] \le 0.
\]
The following result by Wyner~\cite{Wyner1967} (rephrased for our
application) is crucial in proving the above claim:

\begin{lemma}[Uniform covering lemma]
Given $D \in (0,1)$, let $R' > R(D) = (1/2) \log (1/D)$.  Then, for $n
= n(R', D)$ sufficiently large, there exists a dictionary $\Cc_n =
\{\yvh(m): m \in [1:2^{nR'}]\}$ such that for all $\yv$ in the
sphere of radius $r$,
\[
\min_{m \in [1:2^{nR'}]}  ||\yv - \yvh(m)||^2 \le r^2D.
\]
In particular, for all $\yv \in \Real^n$,
\[
\min_{x \in \Real} \min_{m \in [1:2^{nR'}]}
  ||\yv - x \yvh(m)||^2 \le  \|\yv\|^2 D.
\]
%%where the minimum is attained by taking $x = \|\yv\|$.
\end{lemma}

Note that Wyner's uniform covering lemma shows the existence of a
dictionary sequence $\Cc_n$ satisfying
\[
\limsup_{n\to\infty} d_1^{*}(\Cc_n) \le D = 2^{-2R},
\]
which is simply a restatment of the claim for $k = 1$.

Equipped with the lemma, it is straightforward to prove the desired
claim for $k > 1$. Given an arbitrary $\yv$ in the unit sphere,
let $\yvh(m)$ be the best singleton representation of $\yv$
and $\zv_1 = \yv - \yvh(m_1)$ be the resulting error.
Then we find the best singleton representation $\zvh_1 = x_2 \yvh(m_2)$
of $\zv_1$ from the dictionary, resulting in the error
$\zv_2 = \zv_1 - \zvh_1$.
In general, at the $k$-th iteration, the error $\zv_{k-1}$
from the previous stage is represented by $\zvh_{k-1} = x_k
\yvh(m_k)$, resulting in the error $\zv_k$.  Thus this
process gives a $k$-linear representation of $\yv$ as
\begin{align*}
\yv &= \yvh(m_1) + \zv_1 \\
&= \yvh(m_1) + x_2\yvh(m_2) + \zv_2 \\
&= \cdots \\
&= \yvh(m_1) + x_2\yvh(m_2) + \cdots + x_k \yvh(m_k) + \zv_k.
\end{align*}
But by simple induction and the uniform covering lemma, we
have
\[
\|\zv_k\|^2 \le D \|\zv_{k-1}\|^2 \le D^2 \|\zv_{k-2}\|^2 \le D^{k-1}
\|\zv_1\|^2 \le D^k,
\]
which completes the proof of the claim. Note that each of $k$
reprsentations attains mean square error $2^{-2jR_0}$ for its sparsity level
$j = 1, \dots, k$.

\section{Successive Refinement for Gaussian Sources}
The proof in the previous section leads to a deceptively simple proof
of successive refinability of white Gaussian
sources~\cite{Kim2008e}. First note that in the successive linear
representation method we can take $x_k = D^{(k-1)/2}$ for each $k$.
Moreover, if $\Uv = (U_1,\ldots, U_n)$ is drawn independently and
identically according to the standard normal distribution, then it can
be shown that
\[
E\biggl( \frac{1}{\sqrt{n}} \|\Uv\| \;\bigg|\; \frac{1}{\sqrt{n}} \|\Uv\| > 1+\e\biggr)
\cdot \P \biggl( \frac{1}{\sqrt{n}} \|\Uv\| > 1+\e \biggr) \to 0
\]
as $n \to \infty$ for any $\e > 0$. Hence, a good representation of a
random vector $\Uv$ when the vector is inside the sphere of radius
$(1+\e)\sqrt{n}$ is sufficient to a good description of $\Uv$ in general.

Now our successive representation method achieves the (expected) mean square
distortion $(1+\e)D^k$ after $k$ iterations with a dictionary of size
$2^{nR'},$ where $R' > R(D) = (1/2) \log(1/D),$ which is nothing but
the Gaussian rate distortion function. Hence, by describing the index
of the sigleton reprsentation at each iteration using $nR'$ bits, we
can achieve distortion levels $D, D^2, \ldots, D^k$ and trace the
Gaussian rate distortion function for $R', 2R', \ldots, kR'$. (Recall
that we don't need to describe the scaling factors $x_k = D^{(k-1)/2},$
since these are constants independent of $n$.)

More generally, the same argument easily extends to the case in which
incremental rates $R_1, R_2,\ldots, R_k$ are not necessarily
identical; one can even prove the existence of {\em nested} codebooks
(up to scaling) that uniformly cover the unit sphere.

Operationally, the recursive coding scheme for successive refinement
(i.e., describing the error, the error of the error, and so on) can be
viewed as a dual procedure to {\em succesive
cancelation}~\cite{Cover1975c, Wyner1974b} for the Gaussian multiple
access channels, in which the messages for each user is peeled off
iteratively. In both cases, one strives to best solve the single-user
source [channel] coding problem at each stage and progresses
recursively by subtracting off the encoded [decoded] part of the
source [channel output] $\yv$. This duality can be complemented by an
interesting connection between the orthogonal matching pursuit and the
sucessive cancelation~\cite{Jin--Rao2008b} and the duality between
signal recovery and signal representation.  Note, however, that the
duality here is mostly conceptual and cannot be made more precise. For
example, while we can use a single codebook (dictionary) for each of
$k$ successive descriptions (again up to scaling) as shown above, one
cannot use the same codebook for all $k$ users in the Gaussian
multiple access channel. If the channel gains are identical among
users, it is impossible to distinguish who sent which message (from
the same codebook), even without any additive noise! There is no
uniform packing lemma that matches the Gaussian capacity function, to
begin with.

\section{Lower Bound on the Distortion}
We show that for any sequence of dictionaries $\Cc_n$ of size $M = M_n$
and any nondecreasing sequence $k = k_n$,
\begin{align*}
& \log \db_k(\Cc_n) + \frac {2 \log {M\choose k}}{n-k} \\
& \quad + \log \frac{n}{n-k} + \frac{k}{n-k}\log\frac{n}{k} \ge o(1).
\end{align*}
While a similar proof is given in \cite[Theorem 2]{Fletcher--Rangan--Goyal--Ramchandran2006},
we present our version for completeness, which slightly generalizes the proof in
\cite{Fletcher--Rangan--Goyal--Ramchandran2006}.\\
The basic idea of the proof between is to bound the mean square error between
the random vector $\Yv$ and its representation vector $\Yvh$ by computing the
mean square error between $\Yv$ and $\Yvh'$ (a quantized version
of $\Yvh$) and the quantization error (the mean square error between $\Yvh$ and $\Yvh'$).
Then, the tradeoff between the error and the complexity of the representation is
analyzed via rate distortion theory.Details are as follows.

Without loss of generality, assume that
\[
\liminf_{n\to\infty}
\db_k(\Cc_n) \le D < 1.
\]
Let $\yvh = \yvh(\yv) = \sum_{i=1}^k x_i \phiv(m_i)$ be the best
$k$-sparse linear representation of a given vector $\yv$ in the unit
sphere. Then $\yvh$ can be rewritten as
\begin{align}
\yvh = \sum_{i=1}^k \lambda_i(\yv) \psiv_i(\yv),
\end{align}
where $\psiv_1,\ldots,\psiv_k$ form an orthonormal basis of the
subspace spanned by $\phiv(m_1),\ldots, \phiv(m_k)$, uniquely obtained
from the Gram--Schmidt orthogonalization. Since $\|\yvh\|^2 =
\sum_{i=1}^k \lambda_i^2 \le 1$ from the orthogonality of the vectors $\psiv_1,\ldots,\psiv_k$,
$\lambda_i \in [-1,1]$ for all $i$.

We consider two cases:
\begin{enumerate}
\item [(a)] {\em Bounded k}:
Suppose the sequence $k = k_n$ is bounded.
Since $c_n \to 0$ as $n \to \infty$ in Theorem 2 for any bounded sequence $k$,
it is suffices to show that the following inequality holds for any sequence $\Cc_n$
of dictionaries for a bounded sequence $k$.
\[
\frac {2 \log {M\choose k}}{n-k}+ \log\db_k(\Cc_n) \ge o(1).
\]

Next, we approximate $\yvh$ by quantizing
$\lambda_1,\ldots,\lambda_k$ into $\lambda_1',\ldots,\lambda_k' \in
\bigl\{-1, -\frac{l_n -1}{l_n}, \dots, -\frac{1}{l_n}, 0, \frac{1}{l_n}
 \quad ,\frac{2}{l_n},\\ \dots,
\frac{l_n-1}{l_n}, 1 \bigr\}$ with quantization step size $1/l_n$. Let
\begin{align}
\yvh'(\yv) = \sum_{i=1}^k \lambda_i'(\yv) \psiv_i(\yv).
\end{align}
Then,
$\|\yvh - \yvh'\|^2 \le k(1/l_n)^2 = k/l_n^2$
Since $\yvh$ is obtained by orthogonal projection of $\yv$ to the subspace
spanned by $\psiv_1,\ldots,\psiv_k$ and $\yvh'$ is a vector in the subspace,
 $\yv - \yvh$ and  $\yvh - \yvh'$ are orthogonal.
Thus, we have
\begin{align*}
\|\yv - \yvh'\|^2 &= \|\yv - \yvh\|^2 +\|\yvh - \yvh'\|^2 \\
&\le \|\yv - \yvh\|^2 + k/l_n^2
=: d_k(\yv, \Cc_n) + \e_n.
\end{align*}

Now consider a random signal $\Yv$ drawn uniformly from the unit sphere
and its quantized representation
\begin{align}
\Yvh' = \sum_{i=1}^n \lambda_i'(\Yv) \psiv_i(\Yv).
\end{align}
Then, we have $\|\Yv - \Yvh'\|^2 \le \db_k(\Cc_n) + \e_n$.

We have the following chain of
inequalities:
\begin{align}
\notag \log {M\choose k} & + k \log(2l_n +1)\\
\notag &\ge H(m_1(\Yv), \ldots, m_k(\Yv), \lambda_1'(\Yv),\ldots,\lambda_k'(\Yv))\\
\notag &\ge H(\Yvh') \\
\label{eq:rd1} &\ge R(\db_k(\Cc_n) + \e_n),
\end{align}
where
\begin{align*}
R(D) & = \min_{p(\yvh'|\yv): E[ \|\Yv-\Yvh'\|^2] \le \db_k(\Cc_n) + \e_n} I(\Yv; \Yvh')
\end{align*}
is the rate distortion function for $\Yv$ under the mean square distortion $\db_k(\Cc_n) +
\e_n$.

Here are justification for above steps.
The first inequality follows from the ranges of the number of $k$-dimensional
subspaces and $\lambda_j'$. The second inequality follows from the fact that
$\Yvh'$ is a function of $(m_1(\Yv), \ldots, m_k(\Yv), \lambda_1'(\Yv),\ldots,\lambda_k'(\Yv))$.
The last inequality follows from the rate distortion theorem.

By the Shannon lower bound on rate distortion function and the
(Euclidean) volume of the unit sphere,
\begin{align*}
R(D) &\ge h(\Yv) - \frac{n}{2} \log (2\pi e (\db_k(\Cc_n) + \e_n))\\
&\ge \frac{n}{2} \log \left(\frac{1}{\db_k(\Cc_n) + \e_n}\right) - {\log (\pi n)} - \frac{1}{6n}.
\end{align*}
Combined together with \eqref{eq:rd1}, this yields
\begin{align}
\frac{1}{n} & \bigg( {\log {M\choose k}}  + k \log(2l_n +1) \bigg)\\
&\ge \frac{1}{2} \log \left(\frac{1}{\db_k(\Cc_n) + \e_n}\right) - \frac{\log (\pi n)}{n} - \frac{1}{6n^2}\\
&= \frac{1}{2} \log \left(\frac{1}{\db_k(\Cc_n)}\right) + \frac{1}{2} \log \left(\frac{1}{1 + \e_n/\db_k(\Cc_n)}\right)
- o(1).
\end{align}

Now, let $f_n$ be an increasing sequence satisfying
\begin{align}
\lim_{n\to\infty} f_n = \infty \text{\quad and \quad} \lim_{n\to\infty} \frac{\log f_n}{n} = 0,
\end{align}
and take $l_n = f_n (\db_k(\Cc_n))^{-\frac{1}{2}}$.
By plugging $l_n$ to (8), we have
\begin{align*}
& \frac{1}{n} \bigg( \log {M\choose k} + k \log \Big( 2f_n \Big / \sqrt{\db_k(\Cc_n)} +1 \Big) \bigg)\\
&\ge \frac{1}{2} \log \left(\frac{1}{\db_k(\Cc_n)}\right) + \frac{1}{2} \log \left(\frac{1}{1 + \e_n/\db_k(\Cc_n)}\right)
- o(1).
\end{align*}
Arranging the terms in the above inequality yields
\begin{align*}
& \frac{1}{n} \bigg( \log {M\choose k} + k \log \Big( 2f_n + \sqrt{\db_k(\Cc_n)} \Big) \bigg) + o(1)\\
& \ge \frac{n-k}{2n}\log \left(\frac{1}{\db_k(\Cc_n)}\right) + \frac{1}{2}
\log \left(\frac{1}{1 + \e_n/\db_k(\Cc_n)}\right) .
\end{align*}
Then, we can note that $\e_n/\db_k(\Cc_n) = (k/l_n^2)\db_k(\Cc_n) = k/f_n^2$ and
$e_n/\db_k(\Cc_n) \to 0$ as $n \to \infty$. Also, from (9)
$\Big(k \log(2f_n + \sqrt{\db_k(\Cc_n)}\Big)\Big/n \le (k \log(2f_n + 1))/n \to 0$
as $n \to \infty$. Hence, taking the limit $n \to \infty$ to the
last inequality, we get
\[
\liminf_{n\to\infty} \bigg [ \log \db_k(\Cc_n) + \frac {2 k\log M}{n-k} \bigg ] \ge 0.
\]
Finally, it is easy to show that the inequality in Theorem 2 reduces to the
above inequality for the case when $k$ is bounded.

\item [(b)] {\em Unounded k}:
In this case, the scalar quantization in part (a) gives a loose bound. Wyner's uniform covering
lemma, however, can be applied to provide a sharper tradeoff between the description complexity
and the quantization error.

We continue the proof from the orthogonal representation of $\yvh$ in (3).
Since $\yvh$ is a vector with length $\le 1$ in the $k$-dimensional subspace
spanned by $\psiv_1,\ldots,\psiv_k$ and
$k_n$ is an increasing sequence, we can invoke the unform covering lemma.
Therefore, there must exist a dictionary $\Cc'_k$ of size $2^b$
and $\yvh' \in \Cc'_k$ satisfying
\begin{align*}
\| \yvh - \yvh' \|^2  \le 2^{-2b/k}.
\end{align*}
Following the same arguments as in (5)--(9), we have
\begin{align*}
\frac{1}{n} \bigg({\log {M\choose k}}  + b \bigg)
& \ge \frac{1}{2} \log \left(\frac{1}{\db_k(\Cc_n) + 2^{-2b/k}}\right) - o(1).
\end{align*}
Finally, optimizing over b yields
\begin{align*}
\db_k(\Cc_n) \ge 2^{-2 \log {M\choose k}/(n-k)} \cdot \Big( \frac{n-k}{n} \Big) \cdot \Big( \frac{k}{n} \Big)^{k/(n-k)}.
\end{align*}
Taking the logarithm and letting $n \to \infty$ on both sides, we have
the desired inequality.
\end{enumerate}

\iffalse
\section{Concluding Remarks}
The successive representation method presented is geared towards the
information theory friendly regime of exponentially large
dictionaries. Roughly speaking, every signal is sparse (i.e., can be
represented with arbitrarily small distortion by $k \ll n$ reference
signals) in this regime.

What will happen with the more interesting case of subexponential
dictionaries? Theorem 2 is certainly general enough to handle various
regimes of sparsity--dictionary size tradeoffs. But is it also tight
in general, or do we need a separate theory for subexponential
dictionaries?
\fi

\section*{Acknowledgments}
The authors wish to thank Yuzhe Jin and Bhaskar Rao for stimulating
discussions on their formulation of the sparse signal position
recovery problem, which motivated the current work.



% http://www.ctan.org/tex-archive/macros/latex/contrib/supported/IEEEtran/bibtex
%\bibliographystyle{IEEEtran.bst}

\bibliographystyle{IEEEtran}
\bibliography{ref}

\end{document}